\documentclass{article}

\usepackage[nonatbib,final]{wdcs_2020}
    
\usepackage[numbers]{natbib}

\usepackage[utf8]{inputenc} % allow utf-8 input
\usepackage[T1]{fontenc}    % use 8-bit T1 fonts
\usepackage{hyperref}       % hyperlinks
\usepackage{url}            % simple URL typesetting

\usepackage{booktabs}       % professional-quality tables
\usepackage{amsfonts}       % blackboard math symbols
\usepackage{nicefrac}       % compact symbols for 1/2, etc.
\usepackage{microtype}      % microtypography
\usepackage{xspace}
\usepackage{listings}
\usepackage{graphicx}
\usepackage{svg}
\usepackage{wrapfig}

\usepackage{xargs}                      % Use more than one optional parameter in a new commands
\lstdefinestyle{nospaces}{
  showstringspaces=false
}
\title{PrivFramework: A System for Configurable and Automated Privacy Policy Compliance}

\author{
    Usmann Khan \\
    Georgia Institute of Technology \\
    \texttt{ukhan35@gatech.edu} \\
    \And
    Lun Wang \\
    University of California, Berkeley \\
    \texttt{lunwang@berkeley.edu} \\
    \And
    Jithendaraa Subramanian \\
    National Institute of Technology, Tiruchirappalli \\
    \texttt{jithen.subra@gmail.com}
    \And
    Joseph P. Near \\
    University of Vermont \\
    \texttt{jnear@uvm.edu} \\
    \And
    Dawn Song \\
    University of California, Berkeley \\
    \texttt{dawnsong@cs.berkeley.edu} \\
}

\begin{document}
\newcommand{\system}{{\sc PrivFramework}\xspace}
\newcommand{\privguard}{{\sc PrivGuard}\xspace}
\newcommand{\privpolicy}{{\sc PrivPolicy}\xspace}

\maketitle

\begin{abstract}
  Today's massive scale of data collection coupled with recent surges of consumer data leaks has led to increased attention towards data privacy and related risks. Conventional data privacy protection systems focus on reducing custodial risk and lack features empowering data owners. As an end user there are limited options available to specify and enforce one's own privacy preferences over their data. To address these concerns we present \system, a user-configurable framework for automated privacy policy compliance. \system allows data owners to write powerful privacy policies to protect their data and automatically enforces these policies against analysis programs written in Python. Using static analysis \system automatically checks authorized analysis programs for compliance to user-defined policies. 
\end{abstract}

\section{Introduction}
Today our sensitive data is being collected and analyzed by our service providers at a
scale that we are not, and cannot become, aware of. Though we often nominally own our data,
granting others access to it is no simple task. Solutions like OAuth \citep{OAuth} exist to manage data sharing but limit the permission scopes that can be granted to those chosen and implemented by the service provider. Additionally these solutions do not give us control over what happens to our data after permission is granted. For example, Facebook users agreed to share data with app developers but were surprised when in 2018 it came to light that that data had been subsequently shared with political consulting firm Cambridge Analytica \citep{10.1109/MC.2018.3191268}.

In an effort to combat this, governments have introduced regulations to limit the scope of
unauthorized data sharing. These data privacy regulations include the European Union's \emph{General Data Protection Regulation (GDPR)} \citep{EUdataregulations2018}, California's \emph{California Consumer Privacy Act (CCPA)} \citep{CCPA} and the United States' \emph{Health Insurance Portability and Accountability Act (HIPAA)} \citep{HIPAA}.  Unfortunately, today's data processing systems make compliance with these regulations very difficult. This is due in large part to the fact that many existing data processing systems were designed and deployed prior to the conception of these regulations. 

These two problems: the difficulty involved with complying with data regulation and the inability to control our own sensitive data, are deeply interrelated. The root cause is that today's approaches for collecting and processing sensitive personal data require data owners to \emph{give up control} over their data. Once data has left the owner's device, the data owner must \emph{trust} that the organization collecting the data will act in their interest and the organization must go through extreme lengths to ensure their compliance. As can be seen by the numerous recent cases of data misuse \citep{10.1109/MC.2018.3191268}\citep{gressin2017equifax}\citep{affleak}, many organizations have demonstrated that they are not deserving of this trust.

If organizations could ensure verifiable and automatic compliance with privacy policies, it would be safe to trust them with our data. Additionally the ability to specify these policies themselves would extend data owners far greater control and ownership over their data.
\paragraph{\system: returning control to data owners.}
This paper presents \system, an end-to-end framework for building and deploying scalable systems that collect and process sensitive data while enforcing security and privacy policies specified by data owners. \system improves on previous approaches by allowing data owners to specify both \emph{who} may process their data and \emph{how} it may be used.

In \system, data owners submit encrypted data to data collectors, and the data can be decrypted and processed only within \emph{Trusted Execution Environments (TEEs)} that keep the data confidential and ensure the integrity of the results. Data owners also submit policies governing their data, and \system ensures via static analysis that processing tasks performed by the data collector do not violate these policies. These two properties together ensure that data is used in ways that are consistent with its owners' wishes, without requiring trust in the organizations which collect the data.

\paragraph{Related Work}
With the increasing visibility of data privacy issues, there has been increased attention paid to advancing technology in the data privacy space. Our work is most closely related to three directions of privacy research - (1) Data access control \& de-identification, (2) - Pure differential privacy tools, (3) Policy enforcement systems. In Data access control systems there have been advances such as Google DLP \citep{DLP} and Amazon Macie \citep{Macie}. These systems exist to reduce custodial risk but do not allow end users to specify their privacy preferences. Pure differential privacy tools exist in both centralized and decentralized variants and function by adding noise to data. By reducing the signal-to-noise ratio in the data, these solutions often render the data too noisy to provide accurate results. Other privacy policy enforcement systems exist such as those presented by Chowdhury et. al \citep{chowdhury2013privacy} and Sen et. al \citep{sen2014bootstrapping} but these systems do both offer the flexibility of \system while also stopping violations \emph{before they occur}.

\section{The \system System}
\label{sec:overview}
\system provides an end-to-end framework for building configurable privacy-sensitive systems. The framework is comprised of a \emph{client-side API} for building applications that collect data and submit it for analysis, a \emph{collection \& analysis platform} for receiving submitted data, storing it, and executing analyses, and an \emph{analysis API} for writing statistical analyses and machine learning pipelines against the collected data.

Consider the motivating example of a consumer financial application used to track personal budgeting. Currently this requires the use of an interface (e.g. Plaid \citep{plaid} or Yodlee \citep{yodlee}) that requests end users' banking username and passwords, two pieces of extremely sensitive information. Requesting these credentials just to obtain read-only access to transaction history is an overreach that creates a severe and unnecessary privacy hazard. In addition there is no way to limit the scopes of the services' access to data; they have the ability to read and extract any and all information in the account in perpetuity. \system solves these problems by allowing individuals to control the amount of information shared with service providers while simultaneously enabling service providers to provide strong privacy guarantees to reduce custodial risk and build customer trust.

To accomplish this we implement the following workflow:
\begin{enumerate}
    \item The Data Owner uses a client application to upload their sensitive data, along with a formal policy specifying how the data may be processed, to a Data Manager. We use the Oasis Parcel Platform \citep{parcel} to store this data, encrypted with the Owner's key.
    \item The Data Owner uses the Oasis Steward application to authorize Analysts to submit analysis programs to the Data Manager that will be run over the data. 
    \item An authorized Analyst submits an analysis program to the Data Manager. This programs is statically analyzed and constrained according to the specified policy by the Data Manager.
    \item The Data Manager validates the analysis program and returns to the Analyst either the analysis results or a residual policy specifying unsatisfied constraints. At this point the Data Owner has not seen the details of the analysis program, and the analyst has not had direct access to the data. 
    \item If applicable, the Analyst composes a summary or result for the Data Owner.
\end{enumerate}

\begin{figure}[h]
\centering
\includegraphics[width=0.9\textwidth]{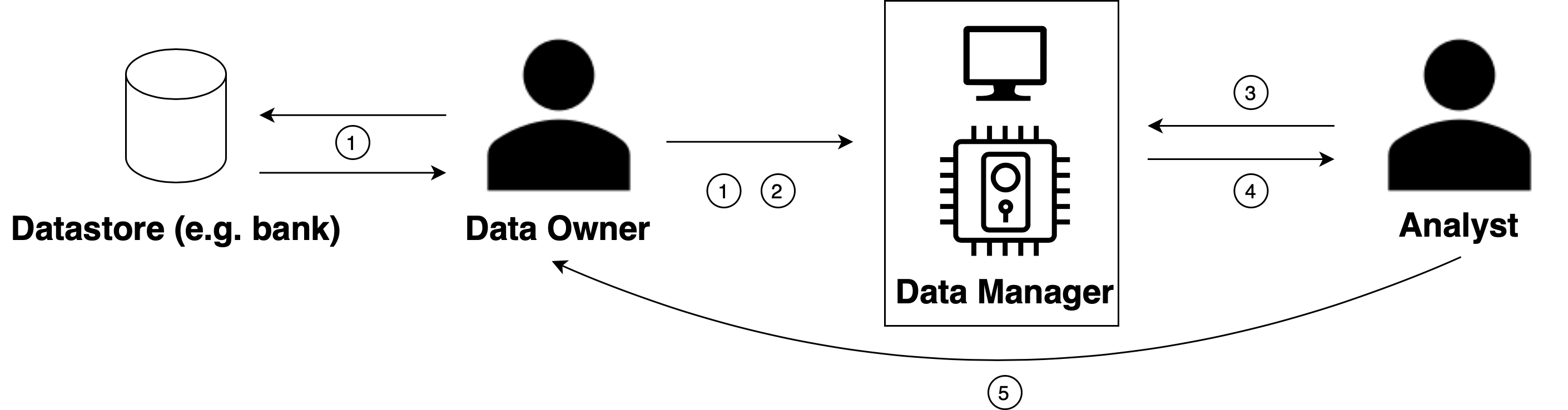}
\caption{Diagram of interactions in the \system system. The circled numbers refer to the above workflow. The Data Manager represents the TEE-enabled system where decryption, compliance checks and analysis occur.}
\end{figure}

Enabling a secure and private workflow like the one described is challenging in several domains. The tool must adhere to strict privacy guarantees, have ergonomic APIs that developers can leverage easily, and provide accessible means for end users to specify their privacy policies. To address these challenges we present \system. In the following sections we present our approach to specifying and enforcing policies, available APIs, the system architecture, and detail a demonstration of the previously described budgeting application designed using \system.

\subsection{Specifying and Enforcing Privacy Policies}
A major challenge in designing privacy sensitive applications is accurately foreseeing the system's eventual use cases and breadth of end user preferences. To address this issue \system is designed to dynamically enforce user specified privacy policies. We have developed a static analysis tool and a Domain Specific Language (DSL) to allow users to configure their own privacy policies, avoiding the need to specify parameters and options in advance. The \system system relies on \privguard, our static analysis framework, and \privpolicy, our DSL, to specify and enforce policies.
\paragraph{Data Capsules.}
Data submitted to \system is stored in \emph{data capsules} \citep{10.1007/978-3-030-33752-0_1}, which pair submitted data with a formal policy governing its use. \system provides analysts with a standard Python environment for processing data capsules, and uses static analysis to ensure compliance with the relevant policies. Our policy analysis supports libraries including Pandas \citep{reback2020pandas}, NumPy \citep{2020NumPy-Array}, TensorFlow \citep{tensorflow}, and PyTorch \citep{pytorch} out of the box. Organizations
can add extend the static analyzer by implementing tracing stubs for their library of choice. 

The data capsule system is \emph{compositional}: processing tasks can be split up into multiple steps that form a processing pipeline. Each program in the pipeline outputs a new data capsule with a new, residual, policy. As programs in the pipeline satisfy policy requirements, those changes are reflected in the new policy attached to the output data capsule. The contents of a data capsule cannot be viewed by the analyst until all of the requirements of its policy have been satisfied.

\paragraph{Policies.} When submitting data to \system, data owners attach policies that govern how the data may be processed. Policies in \system are encoded using the \privpolicy language \citep{10.1007/978-3-030-33752-0_1}, which allows policies to encode both security and access control requirements (e.g. ``only my doctor may view my medical records'') and privacy requirements (e.g. ``my data must be aggregated with data from 100 others before it is processed'' or ``differential privacy must be used when processing my data'').

\paragraph{Static Analysis.} We use \privguard to statically enforce the privacy policies over the Data Owner's data. Only after verifying that the analysis program will not violate any policies is the analysis actually run with access to the data. In the case of noncompliance we also generate a residual policy. This residual policy is packaged with an intermediate analysis state into a new data capsule. If the analysis is compliant then the program's outputs are returned. Otherwise the analyst is provided a reference to the new intermediate capsule. This analysis is done in a TEE to prevent the Data Manager from accessing the data at any time. 

\subsection{Developing Applications with \system}
As a tool meant for developers, all of our components are designed for scalability and interoperability with existing systems. The client-side API is easily integrated with Android applications, and can leverage data stored on the device or drawn from 3rd party sources. The analysis API is implemented as a library for Python programs, so that analysts can re-use existing code and do not need to learn a new programming language. The collection \& analysis platform is built on the Oasis Parcel platform and runs within the AMD SEV \citep{amdsev} Trusted Execution Environment. The collection \& analysis platform is responsible for collecting and processing data, while enforcing data owners' policies.

\paragraph{Client-side Application Development.} \system's client-side API provides application developers with tools for the rapid development of \system
applications. Existing applications can be modified to submit data to \system
in just a few lines of code. The API also provides tools to help data owners
select a policy for submitted data. Applications can leverage the API to allow
data owners to select from a list of possible policies, specify values for
parameterized policies, or even design their own policies from scratch.

\section{Demonstration}

\begin{wrapfigure}{R}{0.4\textwidth} 
\vspace{-10pt}
    \caption{Choosing a policy in the client application is simple for the user and can be customized by the developer.}
    \centering
    \includegraphics[width=0.4\textwidth]{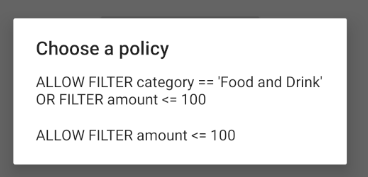}
\end{wrapfigure}

We have implemented a prototype of \system that is capable of demonstrating all of its main features. In this section we will describe the budgeting example discussed in \hyperref[sec:overview]{Section 2}, one of several such applications we have implemented.

We present an application where a Data Owner downloads their own financial data using the APIs that are generally leveraged by service providers. This data is then packaged with a specified policy, uploaded to a Data Manager running a \system agent, and encrypted at rest with their own key. Data Owners then authorize analysts to run analyses over their data. Analysts run these analyses and use the results to generate personal dashboards for the users.

In this process we display the key features of \system:
\begin{enumerate}
    \item The Data Owner privately keeps possession of all secret tokens, passwords, and unencrypted data.
    \item Data Owners may easily specify policies and authorize analysts. 
    \item Analysts submit existing python programs with minimal adjustments to run analyses.
    \item Data Owners do not have access to analysis program details nor do Analysts have direct access to the target data.
\end{enumerate}

\section{Conclusion \& Future Work}
In this paper we have described \system, a new system for user-configurable and automated privacy regulation compliance.\system has the potential to give data owners vastly increased control over their data while also easing organizations' compliance with local data privacy regulations. We implement \system as well as a number of proof of concept applications on top of it, including a personal budgeting system.
In the future we hope to work on further use cases leveraging \system and to extend the list of supported back-end analysis libraries. Future work may also introduce additional advanced privacy features to the policy language such as specialized differential privacy implementations. 
\bibliography{sample}
\end{document}